\documentclass[%
 aip,
 groupedaddress
 amsmath,amssymb,
 preprint,%
]{revtex4-1}

\usepackage{graphicx}% Include figure files
\usepackage{dcolumn}% Align table columns on decimal point
\usepackage{bm}% bold math
\usepackage[utf8]{inputenc}
\usepackage[T1]{fontenc}
\usepackage{mathptmx}
\usepackage{amsmath}
\usepackage{amssymb}

\begin{document}

%\preprint{AIP/123-QED}

\title[]{Energy dependent amplitude of Brillouin oscillations in GaP}
% Force line breaks with \\

\author{Andrey Baydin}
  \email{andrey.baydin@vanderbilt.edu}
\author{Rustam Gatamov}
  \affiliation{Department of Physics and Astronomy, Vanderbilt University, Nashville, TN 37235, USA}
\author{Halina Krzyzanowska}
  \affiliation{Department of Physics and Astronomy, Vanderbilt University, Nashville, TN 37235, USA}
  \affiliation{Maria Curie-Sklodowska University, Pl. M. Cuire-Sklodowskiej 1, 20-031 Lublin, Poland}
\author{Christopher J. Stanton}
  \affiliation{Department of Physics, University of Florida, Gainesville, FL 32611, USA}
\author{Norman Tolk}
    \affiliation{Department of Physics and Astronomy, Vanderbilt University, Nashville, TN 37235, USA}


\begin{abstract}
Gallium phosphide is an important indirect band gap material with variety of applications in optics ranging from LEDs to applications in GaP/Si based solar cells. We investigated GaP using ultrafast, pump-probe coherent acoustic phonon spectroscopy (time-domain Brillouin scattering).  We measured the dependence of the amplitude of the differential reflectivity as modulated by coherent acoustic phonons (CAPs) as a function of laser probe energy and  found that the amplitude of the coherent phonon  oscillations varies non-monotonically  near the direct gap transition at the $\Gamma$ point. A theoretical model is developed which quantitatively explains the experimental data and shows that one can use coherent phonon spectroscopy to provide detailed information about the electronic structure, the dielectric function and optical transitions in indirect band gap materials. Our calculations show that the modelling of experimental results is extremely sensitive to the wavelength dependent dielectric function and its derivatives.
\end{abstract}

\maketitle

\section{\label{sec:Intro}Introduction}
Gallium phosphide (GaP) is a compound semiconductor with an indirect band gap of 2.26 eV \cite{PhysRev.171.876} with a zinc blende crystal structure. GaP is an ideal candidate for optical/photonic structures in the visible range due to its high refractive index and low absorption coefficient \cite{rii}. Most commonly it is used in manufacturing low-cost red, orange, and green light-emitting diodes (LEDs) with low to medium brightness. In addition, GaP is nearly perfectly lattice matched to Si and has a conduction band minimum near the X point like in Si. This allows one to grow high quality layers of GaP on top of Si for possible use in Si-based hybrid optoelectronic devices including high efficiency photovoltaics \cite{Wagner2014, Saive2018}. Recently, the generation of broadband THz pulses by optical rectification in GaP waveguides \cite{doi:10.1063/1.4983371} was demonstrated. The dispersion of the GaP emitter and the peak frequency of the emitted THz radiation are tunable. Also, the use of a waveguide for the THz emission offers scalability to higher power and represents the highest average power for a broadband THz source pumped by fiber lasers \cite{Chang:07}.

Ultrafast laser spectroscopy is a powerful tool for studying the dynamics and characterizing the fundamental interactions  of carriers, spins and phonons in a wide variety of materials.  When one shines an ultrafast, femtosecond pulsed laser beam on a semiconductor surface and observes the decay of the transmission or reflection of a probe pulse as a function of delay time, one can gain valuable insight into carrier relaxation dynamics. Oftentimes, superimposed on the decay signal of the probe pulse is a signal that oscillates in time typically with the period of the $q=0$ optical phonon frequency. These oscillations are known as coherent (optical) phonons. If the absorption of the femtosecond pump pulse is non-uniform (either due to selective absorption in a layered structure or to a short absorption length in a uniform material) then in addition, coherent 
{\it acoustic} phonons (CAPs) can also be generated. The coherent acoustic phonon wavepackets can travel into the sample away from the surface and reflect and scatter from interfaces or structures buried  below the surface. The amplitude of the coherent phonon oscillation (Brillouin oscillations) as a function of delay time can provide information on the quality of surfaces and interfaces as well as internal electric fields. Detection of coherent acoustic phonons is an integral part of the field known as {\it picosecond ultrasonics}. Picosecond ultrasonics including time domain Brillouin scattering is an optical pump probe technique where an ultrafast optical pump pulse generates coherent acoustic phonons (CAP) propagating into a material. The time delayed probe pulse is may then used to detect acoustic echoes coming back to the surface and/or Brillouin oscillations arising from probe light interference \cite{Matsuda2015}. There are many different mechanisms of CAP generation including thermoelasticity, deformation potential, inverse piezoelectric process, electrostriction \cite{Ruello2015, Matsuda2015}. Usually, for efficient CAP wave generation transducer layers made of metals or materials that strongly absorb at the pump energy are used \cite{Miller2006Near-bandgap-wa, Baydin_2016, Baydin2017}. Picosecond ultrasonics is used to measure thin film thicknesses \cite{Wright1992} and elastic properties such as the Young modulus \cite{Chapelon2006}, stress \cite{Dai2016}, sound velocity and index of refraction. In addition, various interactions between photons, excited carriers and phonons are presently under investigation. These include studies of electron diffusion \cite{Tas1994, Wright2001}, shifts in the electronic energy levels by picosecond strain \cite{Akimov2006}, attenuation and dispersion of acoustic phonons \cite{Lin1991, Chen1994}, acoustic solitons and nonlinear acoustics \cite{VanCapel2010, Gusev2014, VanCapel2015}, imaging biological samples \cite{perez2015thin}, adhesion of thin films, two dimensional materials and single cells to a substrate \cite{beardsley2016, Grossmann2017, Ghanem2018}, out of plane energy transfer in Van der Waals materials \cite{Chen2014}, ultrafast acousto-magneto-plasmonics \cite{Temnov2012}, terahertz radiation \cite{Armstrong2009}, and specific mode acoustic phonon - electron interactions \cite{Liao2016}. Analysis of Brillouin oscillations is also employed to study depth dependent optical, acoustical, and acousto-optical parameters of materials \cite{Gusev2018,baydin2019post}.

In this paper, we report on an experimental study of probe energy dependence of the amplitude of Brillouin oscillations arising from CAPs in bulk GaP and compare with theoretical calculations. While normally, one might use CAPs to study the quality of interfaces and anti-phase domains in heterostructures  such as GaP films on Si\cite{ishioka2017sub,ishioka2018coherent}, recently, Ishioka et al. studied the energy dependence of coherent phonons in bulk GaP  with 400 nm (3.1 eV) pump pulses  and probe pulses ranging in energy from 2.0 eV  to 2.6 eV \cite{Ishioka2017}. With a change in energy of the probe pulse, one usually sees a shift in the frequency of oscillation given by: $f= 2 n v /\lambda$ where $n$ is the index of refraction, $\lambda$ the wavelength of the probe and $v$ the sound velocity.   In addition to a change in frequency, Ishioka et al. also surprisingly  saw that  the amplitude of the oscillations increased significantly (by a factor of 5-7) as one approached 2.6 eV. They however, could not quantitatively explain these experimental results.

In this current study, we extend the range of the probe energy from the Ishioka et al.\cite{Ishioka2017} study from below the indirect band gap (2.26 eV) to well  above the direct band gap (2.78 eV) of GaP. Our results show a non-trivial energy dependence (i.e. the change is {\it non-monotonic}) of the amplitude of Brillouin oscillations above 2.6 eV. Our experiments show that the complicated structure in the energy dependence of the amplitude arises from both direct and indirect contributions to the dielectric function of GaP. We develop a theoretical model taking into account the indirect and direct gaps, which shows good agreement with the experimental results, provided we use a experimentally derived dielectric function for GaP {\it with a very small grid size}. We find that dielectric function experimentally obtained by Aspnes\cite{aspnes1983dielectric} and used by Ishioka et al. in their model,  does not have an appropriately small spacing between data points to accurately calculate the derivative in the region of interest.  

\section{Results and Discussion}
\subsection{Amplitude of Brillouin oscillations: Experiment}

A Ti layer (20 nm) was deposited onto a bulk nominally undoped crystal of GaP (100) using an e-beam evaporator with a 2 \AA/s deposition rate. The GaP wafer was purchased from Institute of Electronic Materials Technology, Warsaw, Poland, where it was grown by Liquid Encapsulated Czochralski method. The Ti layer serves as a transducer for efficient generation of the CAP wave. Ti was chosen due to its acoustic impedance that matches one of GaP with only 10\% mismatch and, therefore, acoustic reflections are suppressed at Ti/GaP interface.

Time-domain Brillouin scattering experiments were performed in a standard time-resolved pump-probe setup in reflection geometry. A Coherent Mira 900 with 150-fs pulses at 76 MHz was used as a laser source. Wavelength of the laser was varied. In order to generate the probe wavelength in the UV range, a beta Barium Borate crystal was used. Both beams were focused onto the specimen with spot diameters of 100 $\mu$m and 90 $\mu$m for pump and probe, respectively. The pump beam was chopped using a Thorlabs optical chopper operating at 3 kHz. For energies out of the range of the Coherent Mira 900 (1.38 eV - 1.65 eV, 2.76 eV - 3.3 eV), a Spectra Physics Spitfire Ace amplifier system with 1 kHz repetition rate was used. The pump wavelength of 800 nm (1.55 eV) was chopped at 0.5 kHz. White light generated in a sapphire crystal was used as the probe beam. The probe wavelength was selected using a narrow band pass filter.

\begin{figure}[ht]
    \centering
    \includegraphics[width = 0.9\textwidth]{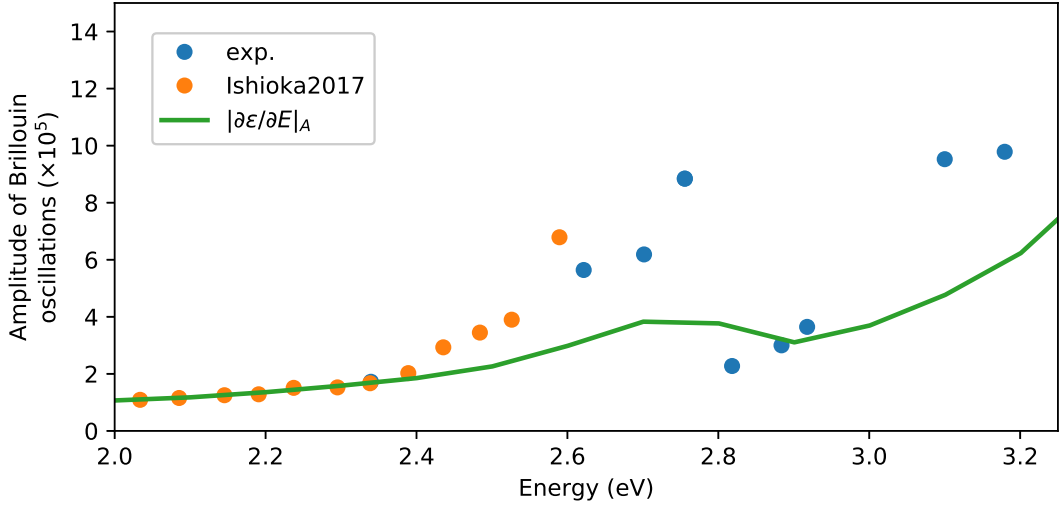}
    \caption{Our experimental results (blue dots) plotted against experimental results (orange dots) obtained by Ishioka et al. and their model (green line)\cite{Ishioka2017} which uses the dielectric function given by Aspnes and Studna\cite{aspnes1983dielectric} on a courser grid.}
    \label{fig:compare}
\end{figure}

Figure \ref{fig:compare} shows our experimental data, the data from Ishioka et al.\cite{Ishioka2017} and the model that utilizes dielectric function from Aspnes et al.\cite{aspnes1983dielectric}. The amplitude increases as the energy increases up to 2.76 eV (direct band gap of GaP). For higher energies past the 2.76 eV the amplitude drops suddenly and starts to increase again. Our data and K. Ishioka et al. data are in a good agreement while the model underestimates the increase in the amplitude of Brillouin oscillations passed 2.4 eV. This is mostl due the fact that the tabulated dielectric function from D. Aspnes's paper does not resolve the feature near $\Gamma$ point and, therefore, leads to a smeared energy derivative of dielectric function. This observation might explain the discrepancy observed by K. Ishioka et al.\cite{Ishioka2017}.

%Figure \ref{fig:fig:exp-model-2}c shows the amplitude of Brillouin oscillations with respect to the probe beam laser energy. The amplitude increases as the energy increases up to 2.76 eV (direct band gap of GaP). Similar behavior for lower energies has been observed previously \cite{Ishioka2017}. For higher energies past the 2.76 eV the amplitude drops suddenly and starts to increase again. 
%\begin{figure}
%    \centering
%    \includegraphics{figures/amplitude_vs_energy_exp.eps}
%    \caption{Experimental amplitude of Brillouin oscillations in GaP. \textbf{Add dashed line to see better}}
%    \label{fig:exp-model}
%\end{figure}
\begin{figure}
    \centering
    \includegraphics[width=\textwidth]{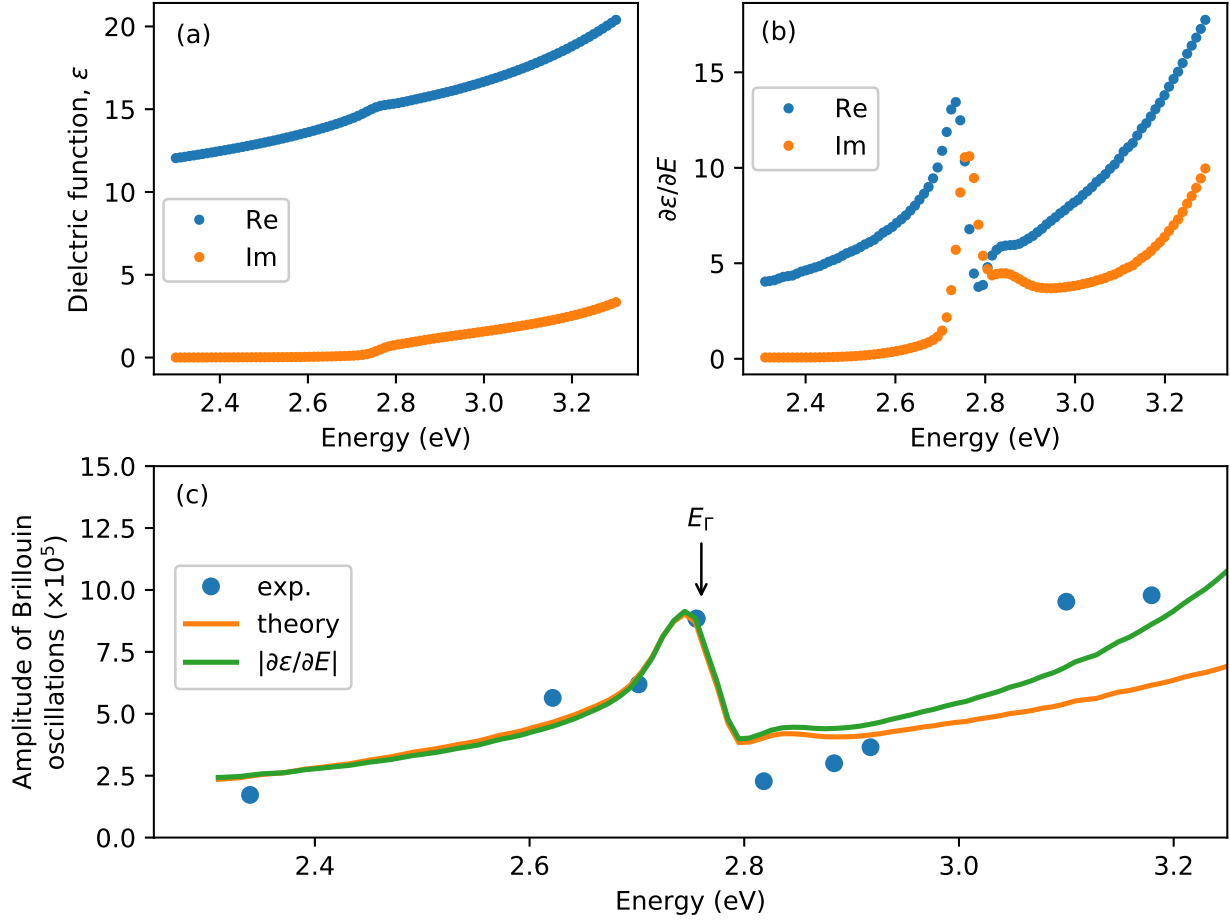}
    \caption{(a) Real (blue) and imaginary (orange) parts of the dielectric function of GaP taken from the database of CompleteEASE software by J. A. Woollam, (b) real and imaginary parts of the energy derivative of dielectric function in Fig. \ref{fig:fig:exp-model-2}a, (c) experimental (blue dots) and theoretical amplitude (orange line) of Brillouin oscillations in GaP versus probe beam energy.}
    \label{fig:fig:exp-model-2}
\end{figure}

\subsection{Amplitude of Brillouin oscillations: Theory}

The amplitude of Brillouin oscillations is derived for a two layer system where the top layer acts as a transducer to generate acoustic pulses. The bottom layer is a substrate, wherein the amplitude of Brillouin oscillations is determined. The complex reflectance of the s-polarized probe in the presence of the travelling CAP wave in a two layer structure with oblique angle of incidence is given by\cite{Matsuda2004}:
\begin{equation}\label{eq:matsuda}
\begin{gathered}
\frac{\delta r}{r}=\frac{ik^{2}}{2k_{0}a_{0}b_{0}}\Bigg[P^{(1)}_{12}\int_{0}^{d}\eta(z',t)\big(a_{1}e^{ik_{1}z'}+b_{1}e^{-ik_{1}z'}\big)^{2}dz'+P_{12}^{(2)}\int_{0}^{\infty}\eta(z'+d,t)\big(a_{2}e^{ik_{2}z'}\big)^{2}dz'\\
+u(0,t)(1-\epsilon_{1})\big(a_{1}+b_{1}\big)^{2}+u(d,t)(\epsilon_{1}-\epsilon_{2})\big(a_{2}\big)^{2}\Bigg],
\end{gathered}
\end{equation}
where $r=b_0/a_0$ is the reflectance for the unperturbed (by the strain wave) sample, $d$ is the thickness of the transducer (top) layer, $k_j = \sqrt{\epsilon_j k^2 - k_x^2}$ is the wave vector in \textit{j}-th medium, \textit{k} is the wave vector in vacuum, $a_j$ and $b_j$ are the electric field amplitudes in \textit{j}-th layer, \textit{u} is the displacement, $\epsilon_1$ and $\epsilon_2$ are complex dielectric functions of the transducer (top layer) and the substrate (bottom layer), respectively. $P_{12}^{(i)}$ is the photoelastic coefficient  for the ith layer \cite{Matsuda2004}. Here, we are interested only in the amplitude of Brillouin oscillations in the substrate, therefore several terms in Equation (\ref{eq:matsuda}) can be omitted. The first term describes photoelastic contribution to the reflectance when the wave is traveling through the transducer layer. Once it leaves the layer, these terms vanish. We ignore any contribution from the static strain caused by the elevated temperature of the transducer layer. Terms that include displacement of the surface and the interface, $u(z,t)=\int_{-\infty}^{z} \eta(z',t) dz'$, also vanish when the strain wave is transmitted into the substrate. Thus, Equation (\ref{eq:matsuda}) can be rewritten as
\begin{equation}\label{eq:drs2}
\frac{\delta r}{r}=\frac{ik^2}{2k_0 a_0 b_0} P^{(2)}_{12} \int_{0}^{\infty} \eta(z'+d,t)a_2^2 e^{2ik_2 z'} dz'.
\end{equation}

In order to determine the amplitude of Brillouin oscillations, we need to know the expression for the generated strain wave, $\eta(z,t)$. To a good degree of approximation, the generated strain pulse can be modeled by a derivative of a Gaussian \cite{Akimov2006}:
\begin{equation}\label{strain-gauss-2}
    \begin{gathered}
        \eta(z,t)=-\eta_0 \frac{(z-vt)}{\xi}\exp\left[-\frac{(z-vt)^{2}}{\xi^{2}}\right],\\
        \eta_0 = \frac{3W\beta B(1-R)(1-r)}{C\xi\rho v_{\text{Ti}}^2},
    \end{gathered}
\end{equation}
where $\xi$ is the absorption depth of the \textit{pump} light, $r$ is the reflection coefficient of longitudinal acoustic waves at the interface between film and substrate, $R$ is the pump light reflection coefficient, $v_{\text{Ti}}$ and $v$ is the sound velocities in transducer layer and substrate, respectively. $W$ is the pump fluence, $\beta$ is the linear expansion coefficient, $B$ is the bulk modulus, $C$ is the volumetric heat capacity, and $\rho$ is the transducer film density. Since $r$ in our experiment is small (0.063), acoustic reflections at the interface were neglected. 

By plugging Equation (\ref{strain-gauss-2}) into Equation (\ref{eq:drs2}), the integral in Equation (\ref{eq:drs2}) can be taken analytically resulting in:
\begin{equation}
    \frac{\delta r}{r_{0}}=-\frac{ik^{2}a_{2}^{2}P_{12}^{(2)}\xi\eta_{0}}{4k_{0}a_{0}b_{0}}\left(e^{-\frac{(d-vt)^{2}}{\xi^{2}}}-i\sqrt{\pi}\xi k_{2}\text{Erfc}\left[\frac{d-vt}{\xi}+i\xi k_{2}\right]e^{-\xi^{2}k_{2}^{2}}e^{2ik_{2}(d-vt)}\right).
    \label{eq:reflectance}
\end{equation}
Brillouin oscillations come from the second term in Equation (\ref{eq:reflectance}). The complementary error function, $\text{Erfc}\left[\frac{d-vt}{\xi}+i\xi k_{2}\right]$, is essentially 2 at longer times. The measured reflectivity change, $\Delta R/R_0$, is related to the complex reflectance change, $\delta r/r_0$, as $\Delta R/R_0 = 2 \text{Re}[\delta r/r_0]$. Therefore, the amplitude of Brillouin oscillations can be expressed as:
\begin{equation}\label{eq:amplitude}
    A_{osc} = \sqrt{\pi} \eta_{0} \xi^{2} \left|\frac{k^{2} k_{2} a_{2}^{2}}{k_{0} a_{0} b_{0}} P^{(2)}_{12} e^{-\xi^{2} k_{2}^{2}} \right|,
\end{equation}

The photoelastic coefficient $P_{12}$ is, in general, a function of energy and defined as \cite{Matsuda2002}:
\begin{equation}\label{photoelastic-dielectric}
    P_{12}=\frac{\partial\epsilon}{\partial\eta}. 
\end{equation}
Since $\epsilon$ is complex, so is $P_{12}$. The strain field, $\eta$, modulates the permittivity, $\epsilon$, by shifting the band gap of a semiconductor by the acoustic deformation potential, $a_{cv}$ \cite{Ishioka2017,PhysRevB.35.7454,PhysRevB.38.12966},
\begin{equation}
    \label{eq:7}
    \epsilon(E,\eta)=\epsilon(E-a_{cv}\eta). 
\end{equation}
The amplitude of strain pulses used TDBS experiments is of the order of $10^{-5}$ and the acoustic deformation potential for most semiconductors is about 10 eV. Therefore, the term $a_{cv}\eta$ is orders of magnitude smaller than $E$. Taking this into account, we can expand Equation (\ref{eq:7}) to the lowest order so that Equation (\ref{photoelastic-dielectric}) becomes
\begin{equation}\label{eq:photoelastic-dielectric2}
    P_{12}=-a_{cv}\frac{\partial\epsilon}{\partial E}. 
\end{equation}
Finally, using Equation (\ref{eq:photoelastic-dielectric2}), the amplitude of Brillouin oscillations from Equation (\ref{eq:amplitude}) takes the following form:
\begin{equation}\label{eq:amplitude2}
    A_{osc} = \sqrt{\pi} \eta_{0} \xi^{2} \left|\frac{k^{2} k_{2} a_{2}^{2}}{k_{0} a_{0} b_{0}} a_{cv}\frac{\partial\epsilon}{\partial E}\bigg|_{E=\hbar\omega} e^{-\xi^{2} k_{2}^{2}} \right|
\end{equation}

The amplitude of Brillouin oscillations in Equation (\ref{eq:amplitude2}) includes the energy derivative of the complex dielectric function. We have taken literature values for the dielectric function of GaP as found in the database of CompleteEASE software by J. A. Woollam, which is shown in Figure \ref{fig:fig:exp-model-2}a. The energy derivative of the dielectric function is taken numerically and shown in Figure \ref{fig:fig:exp-model-2}b. As it can be seen the energy dependence of the derivative already resembles the experimental data shown in Figure \ref{fig:fig:exp-model-2}c. This will be discussed in more detail in the next section. A number of analytical functions have been developed \cite{Dju1999} to fit experimentally obtained dielectric function which can provide smoother energy derivative. However, in our case, the energy range for dielectric function data does not include high energy optical transitions, which may significantly affect the fitting procedure. Finally, the amplitude of Brillouin oscillations is calculated using Equation (\ref{eq:amplitude2}) and compared to the experimental data in Figure \ref{fig:fig:exp-model-2}c. Parameters used in the calculation are reported in Table \ref{table:Ti}.

\begin{table}[ht]
    \centering
    \caption{Parameters for Ti used in the model to calculate strain amplitude, $\eta_0$}
    \label{table:Ti}
    \resizebox{3.37in}{!}{%
    \begin{tabular}{| l l |} 
        \hline
        Physical quantity & Value \\ [0.5ex] 
        \hline\hline
        Absorption depth at 800 nm, $\xi$ & 15.86 nm \cite{Johnson1974}\\ 
        Linear expansion coefficient, $\beta$ & $8.6\times10^{-6} \quad \text{K}^{-1}$ \\
        Bulk modulus, $B$ & 110 GPa \\
        Volumetric heat capacity, $c$ & 2.453$\times10^{-6}$ J/(m$^3$K) \\
        Density, $\rho$ & 4506 kg/m$^3$ \\ %[1ex] 
        Sound velocity, $v_{Ti}$ & 6100 m/s\\
        Reflection coefficient at 800 nm, $R$ & 0.5178\\ 
        Deformation potential, $a_{cv}$ & 10 eV \\[1ex]
        \hline
    \end{tabular}
    }
\end{table}

The amplitude of Brillouin oscillations obtained in the experiment and from the model are in good agreement (see Fig. \ref{fig:fig:exp-model-2}c). There is some discrepancy past the $\Gamma$ point (2.76 eV) of GaP. The model overestimates the amplitude in the energy region between 2.8 eV and 3 eV, and underestimate the amplitude for energies above 3 eV. The results indicate that the amplitude of Brillouin oscillations is maximized near the direct optical transition; at the $\Gamma$ point (2.76 eV). The next direct optical transition is located at $L$ point (zone boundary, $\sim 3.55$ eV) \cite{ref1}, which is out of probe energy range in the current study. Peaking in the amplitude of photoelastic response near $\Gamma$ point has been also observed for GaAs \cite{Miller2006Near-bandgap-wa} and Si \cite{lawler2014experimental}. The dependence of the amplitude of Brillouin oscillations on energy can be explained by multiple optical transitions contributing above the indirect band gap. Our results suggest use of TDBS as a method to measure spectral dependence of the photoelastic response of materials.

It is important to note that the developed model is applicable only for strains $< 10^{-4}$. When CAP waves with larger strain values propagate through a material they shift different valleys by their corresponding deformation potential. In such cases, an analytical model for dielectric function based on critical points \cite{Dju1999} should be employed. Such approach will allow one to find the strain derivative of the dielectric function, $\partial \epsilon/\partial E$, directly by changing critical point values using the energy dependent acoustic deformation potential which varies from valley to valley. 

%Knowledge of the spectral dependence of the photoelastic response is important both from fundamental point of view and for applications. Particularly, it is relevant to nanoscale imaging using TDBS \cite{Gusev2018}, where one would want to optimize and choose a probe wavelength leading to higher amplitude of Brillouin oscillations and, consequently, larger signal-to-noise ratio.  For example, in work on coherent acoustic phonons in InMnAs films on an GaSb substrate, it was seen that the large magnitude of the Brillouin oscillations resulted from the fact that the Brillouin oscillations were probed near peaks in the derivatives of the complex dielectric function \cite{Wang2005, Sanders2005}.

\subsection{Simple Model}
Insight is gained if one studies a much simplified model of transient reflectivity. It can be shown that if the changes $\delta r$ to the reflectivity due to the pump pulse are small, $\delta r << r_0$, then
\begin{equation}
    \label{ap10d}
    \frac{\Delta R}{R_0}= 2\text{Re}{\left(\frac{\delta r}{r_o}\right)}.
\end{equation}
For a uniform material with small changes to the index of refraction $\delta n$ induced by a pump pulse, then it can be shown\cite{cook2010ultrafast}:
\begin{equation}\label{ap14}
    \frac{\Delta R}{R_0}\propto 2\text{Re}\left\{\int_{0}^{\infty}e^{i2kx}\frac{\delta}{\delta
    x}\delta n(x,t) dx\right\}.
\end{equation}
where $n$ is the complex index of refraction, $k = n k_0$ and $\delta n$ is the change in the complex index of refraction due to the pump pulse. For GaP, with laser energies between 2 eV and 3.5 eV, the real part of the index of refraction is much larger than the imaginary piece and can therefore be treated as real. Note however, that changes to the index of refraction due to the pump pulse can be either real or imaginary.

The change in the index of refraction is related to the Seraphin coefficients $\partial \epsilon / \partial E$ through 
\begin{equation}\label{ap15}
2 n \delta n=\delta \epsilon=-a_{cv}\frac{\partial \epsilon}{\partial E} \eta.
\end{equation}
As a result, we obtain:
\begin{equation}\label{ap16}
\frac{\Delta R}{R_0}\propto 2\text{Re}\left\{ \frac{\partial\epsilon}{\partial E}\int_{0}^{\infty}e^{i2kz}\frac{\delta}{\delta
z}\delta\eta(z,t) dz\right\}.
\end{equation}
Since the strain wave propagates undistorted through the medium, $\eta(z,t)=\eta_0(z-vt)$ where $\eta_0$ is the initial strain. We can change variables to $u=z-vt$ and then for long times (so all the strain is propagating in the positive direction away from the surface) that  
\begin{equation}\label{ap17}
\frac{\Delta R}{R_0}\propto 2\text{Re}\left\{ \frac{\partial \epsilon}{\partial E}~~e^{i2kvt}\int_{-\infty}^{\infty}e^{i2ku}\frac{\delta}{\delta
u}\delta\eta_0(u) du\right\}.
\end{equation}
The last term (integral) is a form factor which depends only on the initial strain profile which depends on the pump pulse parameters. The middle term is the oscillatory part of the reflection. This has a dependence on the frequency of the probe laser light (which determines the period of oscillations) but does not influence the amplitude of the oscillations. Only ${\partial\epsilon/\partial E}$ affects the amplitude of the oscillations for different probe energies. We can write
\begin{equation}
   \frac{\partial\epsilon}{\partial E}=\sqrt{\left(\frac{\partial\epsilon_r}{\partial E}\right)^2
   +\left(\frac{\partial\epsilon_i}{\partial E}\right)^2} ~~e^{i\phi_{\epsilon}}
\end{equation}
so we see that the dominant contribution to the amplitude of the signal is given by 
\begin{equation}
 A_{osc}\propto \left| \frac{\partial\epsilon}{\partial E}\right|=\sqrt{\left(\frac{\partial\epsilon_r}{\partial E}\right)^2
   +\left(\frac{\partial\epsilon_i}{\partial E}\right)^2}.
\end{equation}
This is plotted by the green line in Figure \ref{fig:fig:exp-model-2}c. As can be seen from the Figure \ref{fig:fig:exp-model-2}c, the green curve agrees remarkably well with both the experimental data (blue dots) and the more complete theory (orange curve).

%%%%%%%%%%%%%%%%%%%%%%%%%%%%%%%%%%%%%%%%%%%%%%%%%%%%%
\section{Conclusion}

Knowledge of the spectral dependence of the photoelastic response is important both from fundamental point of view and for applications. Particularly, it is relevant to nanoscale imaging using TDBS \cite{Gusev2018}, where one would want to optimize and choose a probe wavelength leading to higher amplitude of Brillouin oscillations and, consequently, larger signal-to-noise ratio.  For example, in work on coherent acoustic phonons in InMnAs films on an GaSb substrate, it was seen that the large magnitude of the Brillouin oscillations resulted from the fact that the Brillouin oscillations were probed near peaks in the derivatives of the complex dielectric function \cite{Wang2005, Sanders2005}.

In conclusion, we have investigated the photoelastic response of GaP as probed by time-domain Brillouin scattering. The results show order of magnitude changes in the amplitude of Brillouin oscillations with respect to probe energy which is maximized near direct optical transitions. Calculations based on the developed theoretical model are in a good agreement with experimental data. The results obtained in this paper are of importance to the understanding of detection mechanisms of coherent acoustic phonons in indirect band gap semiconductors and GaP based opto-electronic devices. Information obtained from these types of studies can be used for optimizing the optical response for a wide variety of materials.

\begin{acknowledgments}
The authors acknowledge the ARO for financial support under Award No. W911NF14-1-0290. Portions of this work were completed using the shared resources of the Vanderbilt Institute of Nanoscale Science and Engineering (VINSE) core laboratories. CJS was supported by the Air Force Office of Scientific Research under Award No. FA9550-17-1-0341.
\end{acknowledgments}

%\appendix
%\nocite{*}
%%%%%%%%%%%%%%%%%%%%%%%%%%%%%%%%%%%%%%%%%%%%%%%%%%%%%
\bibliography{references_list}

\end{document}